\documentclass[onecolumn,aps,amsmath,amssymb,showpacs,amsfonts,floatfix,superscriptaddress]{revtex4}
\usepackage{graphicx}% Include figure files
\usepackage{dcolumn}% Align table columns on decimal point
\usepackage{bm}% bold math
\usepackage[hypertex]{hyperref}
\usepackage{latexsym}
\usepackage{float}
\usepackage{supertabular}
\usepackage{longtable}

\begin{document}
\title{Dynamics of Strategic Three-Choice Voting}
\author{D. Volovik}
\affiliation{Center for Polymer Studies and Department of Physics, Boston University, Boston,
Massachusetts 02215 USA}
\author{M. Mobilia}
\affiliation{Department of Applied Mathematics, School of Mathematics,
  University of Leeds, Leeds LS2 9JT, U.K.\footnote{Permanent address from
    January 2009.  Address from January 2009: M.Mobilia@leeds.ac.uk.}}
\affiliation{Mathematics Institute \& Warwick Centre for Complexity Science,
The University of Warwick, Coventry CV4 7AL, U.K.}
\author{S. Redner}
\affiliation{Center for Polymer Studies and Department of Physics, Boston University, Boston,
Massachusetts 02215 USA}
\date{\today}

\begin{abstract}
  
  In certain parliamentary democracies, there are two major parties that move
  in and out of power every few elections, and a third minority party that
  essentially never governs.  We present a simple model to account for this
  phenomenon, in which minority party supporters sometimes vote ideologically
  (for their party) and sometimes strategically (against the party they like
  the least).  The competition between these disparate tendencies reproduces
  the empirical observation of two parties that frequently exchange majority
  status and a third party that is almost always in the minority.

\end{abstract}


\maketitle

\section{Introduction}

A feature of governance in several countries with parliamentary
elections---typically British Commonwealth countries and Britain itself---is
that two major parties move in and out of power every few elections, while a
smaller third party either has never or rarely governed.  This lack of
representation of the minority party occurs even though its vote fraction can
be close to that of the major parties.  

Governance is determined by the party (or coalition) that has the majority of
members of parliament (MPs).  Each MP is the candidate with the most votes in
each parliamentary district election.  This voting system makes it difficult
for a minor party to gain representation that mirrors its vote fraction.  For
example, if one party (out of three) receives 30\% of the vote in every
district while the other two parties equally share the remaining 70\%, then
the minority party will have no MPs although it is supported by nearly 1 in 3
voters.  As an illustration \cite{british-elections}, in the 1983 British
election, the Conservative party won 42.4\% of the popular vote and 397 seats
(61.1\% of 650 seats), the Labor party won 27.6\% of the vote and 209 seats
(32.2\% of 650), while the traditional third-place Liberal party won 25.4\%
of the vote but only 23 seats (3.56\% of 650).  Similarly, Canada had two
major parties for much of the $20^{\rm th}$ century---Liberals (center-left),
Progressive Conservatives (center-right), and a smaller, but still national
scale, New Democratic Party (leftist) that has never governed
\cite{canadian-elections,sc}.

\begin{figure}[ht]
\centering
\includegraphics*[width=0.31\textwidth]{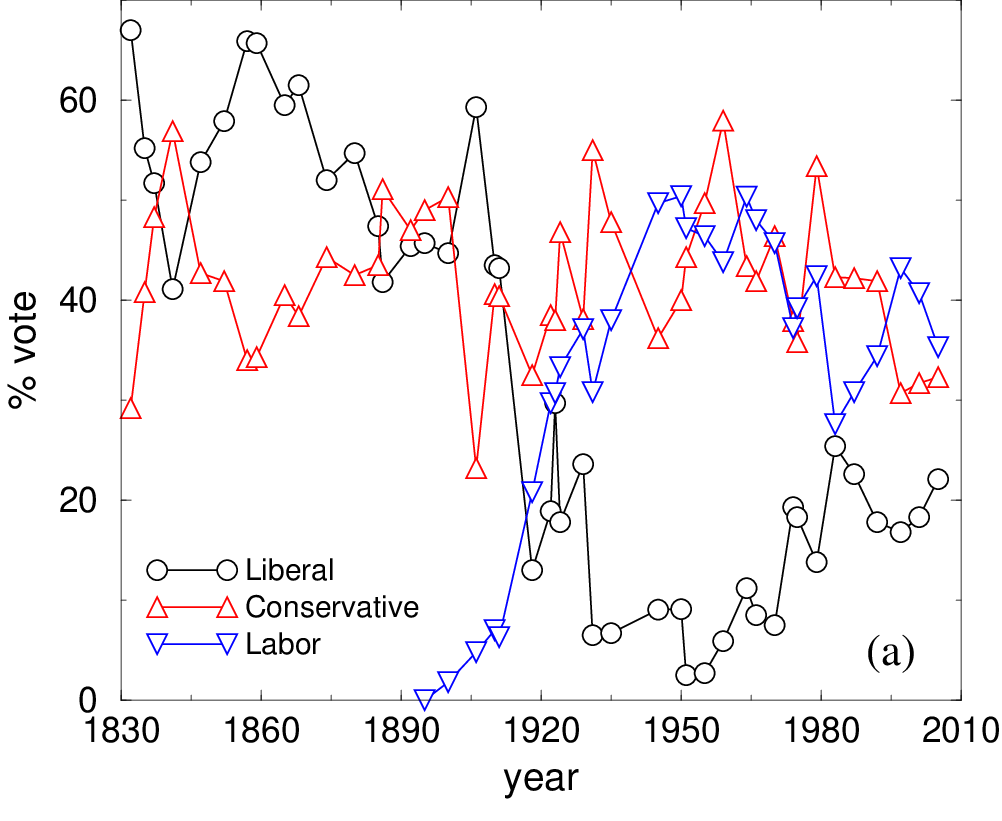}\hskip 0.2in
\includegraphics*[width=0.295\textwidth]{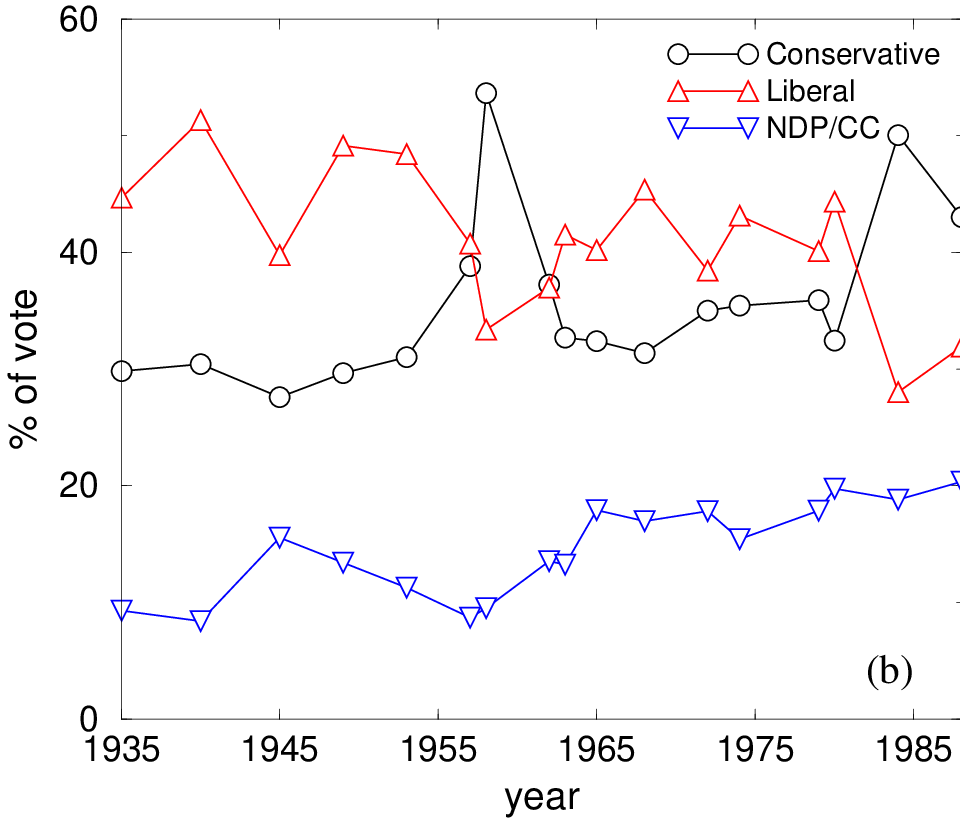}\hskip 0.2in
\includegraphics*[width=0.31\textwidth]{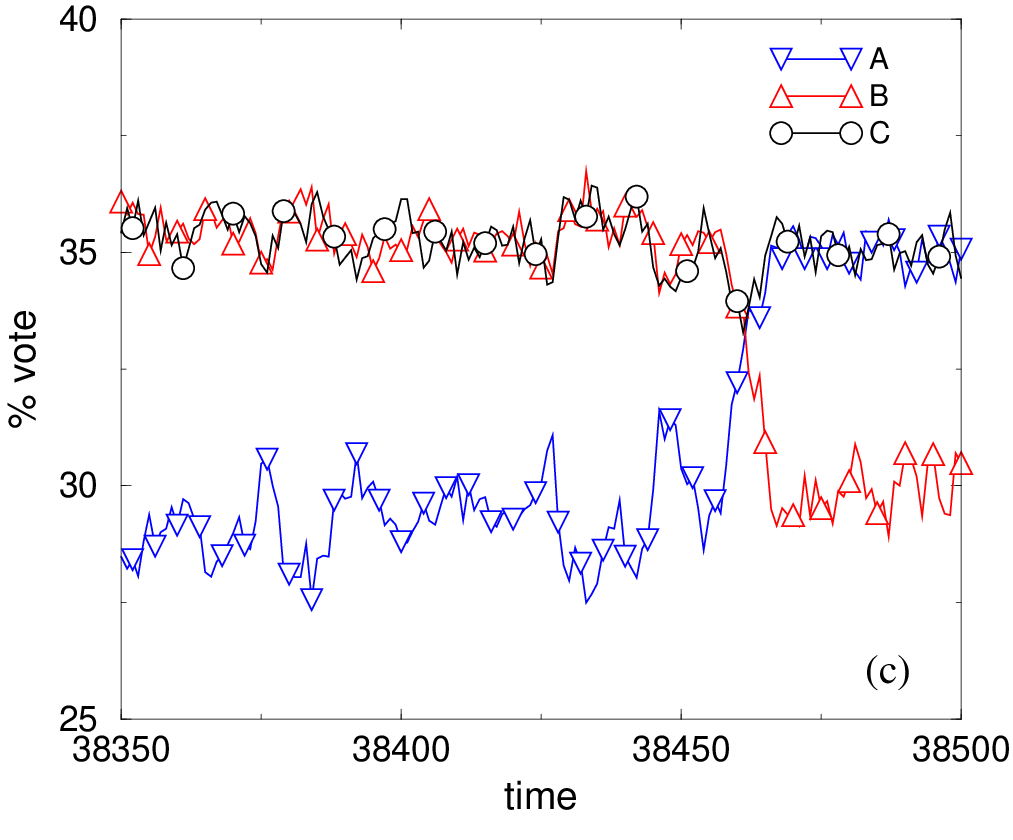}
\caption{(a) British election results since 1830. (b) Canadian election
  results from 1935--90; here NDP/CC denotes the New Democratic Party and its
  Cooperative Commonwealth party antecedent. (c) Representative simulation
  results for the time evolution of the densities of the three states in the
  strategic voter model with state-dependent strategic bias for $50000$
  agents and parameter $x_0=0.104$ (see text for definition of $x_0$).}
  \label{elections}
\end{figure}

To understand how such a voting pattern can arise, consider the case of the
traditional minority Liberal party in Britain.  If their election fortunes
seem promising, then supporters are likely to be galvanized to vote for their
party.  However, if Liberal prospects seem bleak and the diametrically
opposed Conservative party appears strong, a Liberal may vote for the Labor
party to forestall the Conservatives.  This strategy clearly disfavors the
minority party (Fig.~\ref{elections}).  In fact, when the number of parties
is larger than two, ambiguous outcomes for the voting preference, such as the
Condorcet paradox \cite{cond}, can easily arise.  There is also a richer
range of phenomenology than in two-party voting \cite{galam}.

To account for the endemic weakness of a minority party, we introduce the
``strategic'' voter model that encapsulates the strategic/ideological
dichotomy outlined above.  The original voter model~\cite{voter} is a
paradigmatic non-equilibrium process that describes ordering in
non-equilibrium systems and consensus in opinion
dynamics~\cite{Castellano_review}.  In the voter model, agents are endowed
with a 2-state opinion variable.  The dynamics is defined by picking a voter
at random and updating its opinion and repeating {\it ad infinitum}.  In the
update event, the agent adopts the opinion of a random neighbor.  Neighbors
can be defined on a complete graph (interaction with any other agent
equiprobably), across the links of regular lattices~\cite{VMlattice} or
complex networks~\cite{VMnets}, or on adaptive graphs~\cite{adaptive}.
Various extensions of the dynamics itself have been considered
\cite{extensions}.

In the next section, we define two natural versions of strategic voting and
determine their dynamics in the mean-field limit.  Typically, the population
is driven to a steady state where a single party is perpetually in the
minority unless the strategic bias is extremely weak.  In the
symmetric-breaking steady state, discrete fluctuations ultimately cause the
minority party to escape its minority status on a slow time scale.  It
appears possible that election results (Fig.~\ref{elections}) would exhibit
these same oscillations if the data continued for a much longer time period.

\section{Strategic Voting}

We consider a population of $N$ voters on a complete graph that have three
possible opinions states: $A$, $B$ and $C$.  The idealized complete graph is
used because of its simplicity and analytical tractability.  Each of these
states is equivalent, in contrast to real political parties, and our model
thus does not incorporate any subjective political attributes.  We denote the
fractions of the three types of voters as $a$, $b$, and $c$.  The state of
each individual evolves by the following voter-like dynamics:
\begin{itemize}
  
\item With rate $T$, a voter spontaneously changes to another state
  equiprobably.  This thermalization step ensures that consensus is never
  reached.
  
\item With rate $r_{ij}$, a random pair of voters in states $i$ and $j$ is
  picked and one member of this pair adopts the state of the other voter.  In
  general, $r_{ij}$ does not equal $r_{ji}$ (see below).
\end{itemize}

In the mean-field limit ($N\to \infty$), the fractions of each voter species evolve according
to the rate equations:
\begin{align}
\label{RE}
\dot a &= T(b+c-2a) + r_{AC}\, ac + r_{AB}\, ab \nonumber \\
\dot b &= T(c+a-2b) + r_{BA}\, ba + r_{BC}\, bc  \\
\dot c &= T(a+b-2c) + r_{CA}\, ca + r_{CB}\, cb \,.\nonumber
\end{align}
The terms proportional to $T$ account for spontaneous opinion changes, while
the remaining terms account for voter-like updating.  In the classic voter
model \cite{voter,Redner}, each $r_{\alpha\beta}=0$, because a voter pair
$\alpha\beta$ can change to $\alpha\alpha$ or to $\beta\beta$ equiprobably;
thus the density of each species is conserved.  In our model, we only require
$r_{\alpha\beta}+r_{\beta\alpha}=0$ to conserve the total density.

For the strategic voter model, we take $r_{\alpha\beta}$ to be nonzero when
one of $\alpha$ or $\beta$ denotes a minority state.  A simple choice is a
strategic bias that is independent of the fractions of each species, {\it
  viz}:
\begin{equation*}
\label{rab}
r_{AB}=-r_{BA} = \begin{cases}
+r & B~ {\rm minority},\\ ~~0 & C~ {\rm minority}, \\ -r & A~ {\rm minority},
\end{cases}
\end{equation*}
(and cyclic permutations for $r_{AC}$ and $r_{BC}$).  Thus if $B$ is in the
minority, then an $AB$ interaction favors the outcome $AA$ rather than $BB$
because $r_{AB}>0$ and $r_{BA}<0$.  Conversely, if $A$ is in the minority, an
$AB$ interaction favors $BB$ rather than $AA$.  If neither $A$ or $B$ are in
the minority, then they undergo conventional voter dynamics in which their
average densities do not change.  While it would be politically more
realistic to have non-symmetric interactions in which the minority species
has a definite preference for one of the two non-minority species, this more
complicated model does not yield any additional insights about strategic
voting.

In the $c_<$ sector of the composition triangle (where $c$ is in the
minority, see Fig.~\ref{triangle}), the rate equations reduce to:
\begin{eqnarray}
\label{REc}
\dot a &=& T(1-3a) + r ac  \nonumber \\
\dot b &=& T(1-3b) + r bc \\
\dot c &=& T(1-3c) - rc(1-c) . \nonumber
\end{eqnarray}
Here we use $a+b+c=1$ to simplify the $T$-dependent terms and the last term
in the equation for $\dot c$.  Similar equations hold for the $a_<$ and $b_<$
sectors by cyclic permutations.

We will also study a more realistic strategic voting in which the strategic
bias vanishes as the minority population approaches those of the other two
states.  In terms of election sentiment, if a minority party supporter
believes that his/her party will win in an upcoming election, then there is
every reason to vote ideologically with the party and not strategically.  For
simplicity we consider the case where the strategic bias varies linearly in
the depth of minority status as quantified by the density-dependent strategic
voting rate $r=r_0[(a+b)/2-c]$ in Eqs.~\eqref{REc} (the $c_{<}$ sector), and
analogously for the other sectors of the composition triangle.

\section{Solution to the Rate Equations}

\subsection{Constant Strategic Bias}

\begin{figure}[ht]
\centerline{\includegraphics*[width=0.335\textwidth]{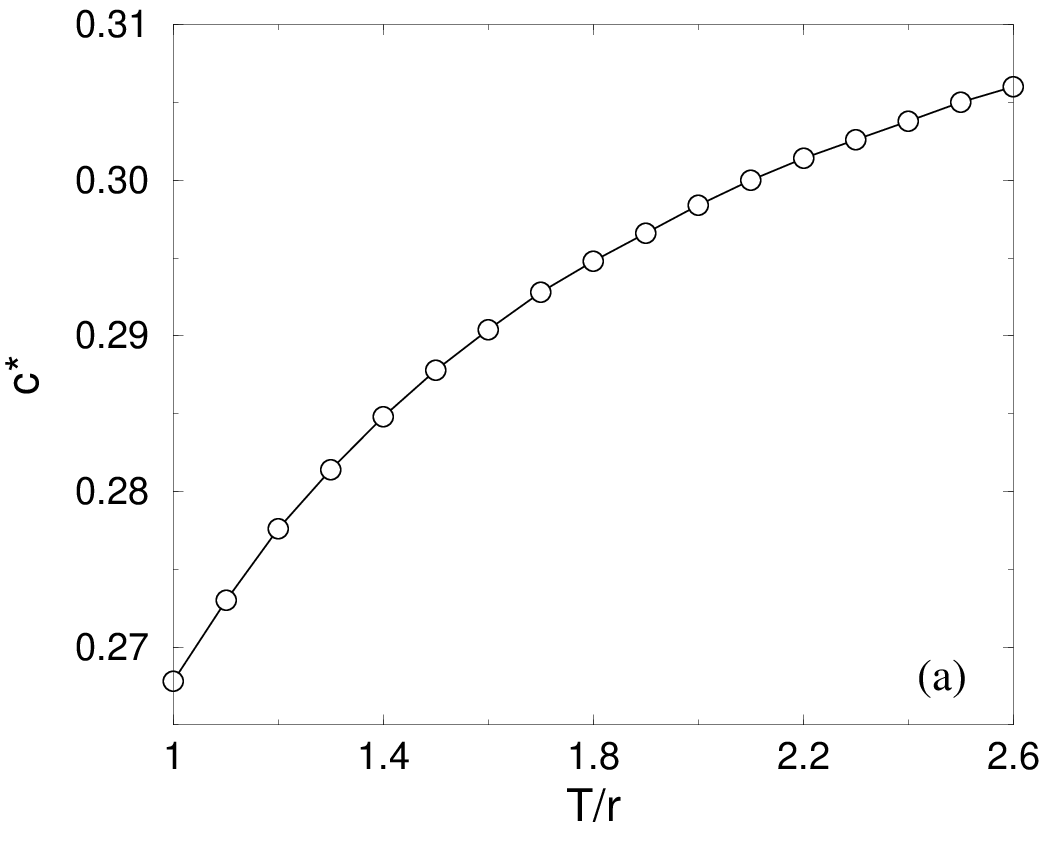}
\hskip 0.75in \includegraphics*[width=0.335\textwidth]{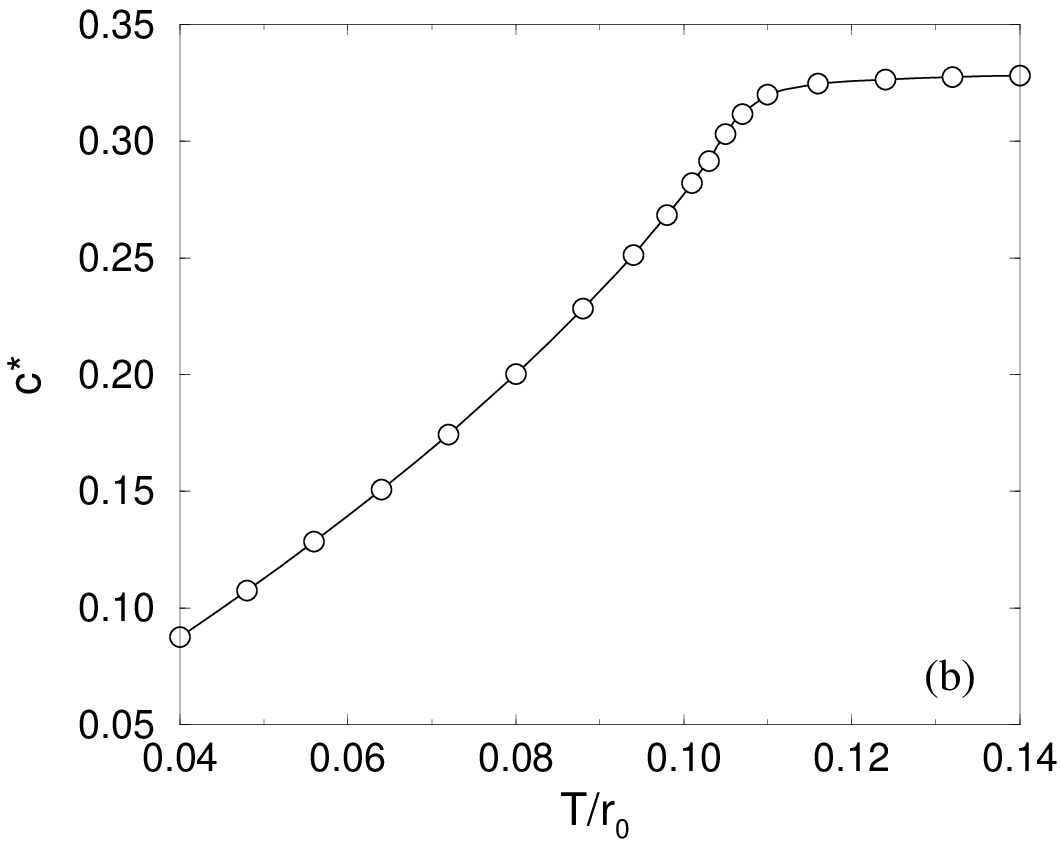}}
\caption{Simulation results for the fixed point minority concentration $c^*$
  for: (a) a population of $N=5000$ agents with constant strategic bias as a
  function of $x=\frac{T}{r}$, and (b) $N=50000$ agents with state-dependent
  strategic bias as a function of $x_0=\frac{T}{r_0}$.  The numerical value
  $c^*$, which agrees perfectly with the rate equation predictions, is
  determined by the time integral of the concentration divided by the
  observation time.  The change in behavior when $x_0\approx \frac{1}{9}$
  becomes a cusp as $N\to\infty$.}
  \label{c-vs-N}
\end{figure}

For a strategic bias with a fixed rate $r$, we solve the last of
Eqs.~\eqref{REc} by rewriting it in the factorized form
\begin{equation}
\label{ct}
\frac{1}{r}\frac{dc}{dt} = c^2-c\left(1+{3x}\right)+x \equiv (c-c_+)(c-c_-).
\end{equation}
Here $x\equiv\frac{T}{r}$ quantifies the relative importance of the strategic
bias (with strong bias corresponding to small $x$), and
\begin{equation*}
c_\pm\,\,= \frac{1}{2}\left[(1+3x) \pm 
\sqrt{1+2x+9x^2}\right]\equiv 
\frac{1}{2}\left[(1+3x)\pm\frac{1}{\tau}\right].
\end{equation*}
We now apply a partial fraction expansion to Eq.~\eqref{ct} to render it
integrable by elementary means and the solution is
\begin{eqnarray}
\label{app-c}
c(t)= \frac{c_--c_+\,\mathcal{C}\, e^{-rt/\tau}}{1-\mathcal{C}\,e^{-rt/\tau}},
\qquad {\rm where}\qquad \mathcal{C}= \frac{c(0)-c_-}{c(0) - c_+}~.
\end{eqnarray}
Thus $c(t)$ converges to its steady-state value of $c^*=c_-$ exponentially in
time.  The value of $c^*$ sharply goes to zero as $x\to 0$ because the
strategic bias dominates, while $c^*$ gradually approaches the limiting value
of $\frac{1}{3}$ when $x\to \infty$ (see Fig.~\ref{c-vs-N}(a)).

\begin{figure}[ht]
\centerline{\includegraphics*[width=0.25\textwidth]{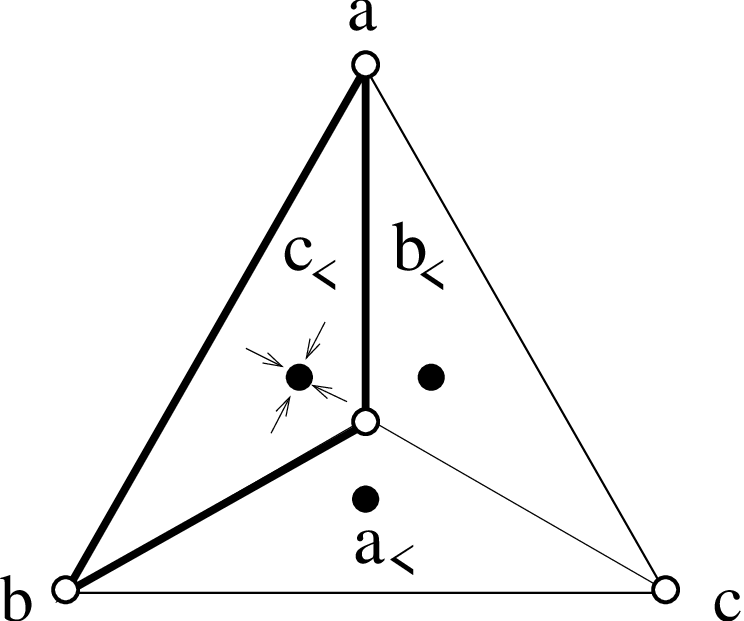}}
\caption{Composition triangle showing the stable (dots) and unstable
  (circles) fixed points of the rate equations.  The triangle represents the
  locus of points $a+b+c=1$ in the $abc$ density space.  Each sector is
  demarcated by separatrices, and the local flow near one stable fixed point
  is shown.  The heavy lines outline the $c_<$ sector where the density of
  $c$ is in the minority.}
  \label{triangle}
\end{figure}

For the majority density, we write the first of Eqs.~\eqref{REc} as $\dot a
+a f =T$, where $f(t)\equiv 3T-rc(t)$, with formal solution
\begin{equation}
\label{a-formal}
a(t) = T\,e^{-F(t)}\int_0^t e^{F(t')}\,dt' +a(0)\, e^{-F(t)}~,
\end{equation}
where 
\begin{equation*}
F(t) =\int_0^t f(t')\, dt
=3Tt-r\int_0^t  \frac{c_--c_+\mathcal{C}\, e^{-rt'/\tau}}{1-\mathcal{C}\,e^{-rt'/\tau}}\, dt'
= -\ln\left[\frac{1-\mathcal{C}}{1-\mathcal{C}\,e^{-rt/\tau}}\right] 
+\frac{rt}{2\tau}[1-\tau(3x-1)].
\end{equation*}
Finally, we substitute the above result for $F(t)$ into Eq.~\eqref{a-formal}
and perform the integral to obtain
\begin{eqnarray}
\label{a}
a(t) = {a^*} - \frac{ e^{-rt/\tau}}{1+\eta}  
\left[\frac{\mathcal{C}(1-\mathcal{C})}{1-\mathcal{C}e^{-rt/\tau}}\right]\, 
+\left[a(0) -\frac{2x\tau\left\{1+\mathcal{C}+\eta(1-\mathcal{C})\right\}}
{(1-\mathcal{C})\left\{1-\eta^2\right\}}\right] 
\left[ \frac{(1-\mathcal{C}) e^{-rt(1-\eta)/2\tau}}{1-\mathcal{C} e^{-rt/\tau}}\right] ~,
\end{eqnarray}
and similarly for $b(t)$.  Here $\eta\equiv (1-3x)\tau$, and the steady-state
densities are $a^*=b^*=(1-c^*)/2$.  It is easy to check that $0<1-\eta<2$, so
that the approach to the steady-state is dominated by $a(t)-a^* \sim
e^{-rt(1-\eta)/2\tau}$ in Eq.~(\ref{a}), while the minority species relaxes
more quickly to stationarity: $c(t)-c^* \sim e^{-rt/\tau}$.  The same
considerations apply, {\it mutatis mutandis}, in the $a_{<}$ and $b_{<}$
sectors of the composition triangle.

The non-trivial fixed points (dots in Fig.~\ref{triangle}) are the basins of
attractions for the $a_<$, $b_<$ and $c_<$ sectors for {\em any} value of the
basic parameter $x$, and for almost all initial conditions (except for the
special cases where all the densities initially equal or only two densities
are nonzero).

\subsection{State-Dependent Strategic Bias}

We now consider strategic bias with state-dependent rate $r=r_0[(a+b)/2-c]$
in the rate equations \eqref{REc}.  To simplify technical details, we assume
$a=b$ (corresponding to an initial state $a(0)=b(0)$) throughout the
evolution; thus the normalization condition now is $2a+c=1$.  The rate
equation for the minority density $c$ in the $c_<$ sector now is:
\begin{equation}
\label{c-var}
\dot c = (1-3c)T-\frac{r_0}{2}c(1-c)(1-3c)\equiv -\frac{3r_0}{2}(c-c_-)(c-c_+)(c-c_3),
\end{equation}
where $c_3=\frac{1}{3}$, $c_\pm= \frac{1}{2}\left(1 \pm\sqrt{1-8x_0}\right)$,
and $x_0\equiv\frac{T}{r_0}$.  By plotting the right-hand side of
Eq.~\eqref{c-var} as a function of $c$, it is clear that the stable fixed
point value is $c^*=c_-$ for $x_0<\frac{1}{9}$.  In this case, any initial
state in the $c_<$ sector flows to the fixed point
$(\frac{1-c^*}{2},\frac{1-c^*}{2},c^*)$; this behavior is qualitatively
similar to that of constant strategic bias.  Conversely, for
$x_0>\frac{1}{9}$ the stable fixed point is $c^*=\frac{1}{3}$, with $c_+$ and
$c_-$ both greater than $\frac{1}{3}$ (Fig.~\ref{c-vs-N}, right).  Thus a
sufficiently weak strategic bias or a sufficiently strong voting uncertainty,
as quantified by $x_0>\frac{1}{9}$, leads to the fixed point
$(\frac{1}{3},\frac{1}{3},\frac{1}{3})$ becoming stable.  This change in the
stability of the fixed points as a function of the basic parameter $x_0$ is
an unexpected feature of state-dependent strategic voting.

To solve for the time dependence, we perform a partial fraction expansion to
transform the rate equation \eqref{c-var} to
\begin{equation*}
dc\,\left[\frac{1}{c_3-c_-}\left(\frac{1}{c-c_3}-\frac{1}{c-c_-}\right)-
\frac{1}{c_3-c_+}\left(\frac{1}{c-c_3}-\frac{1}{c-c_+}\right)\right] = 
\frac{3}{2}(c_+-c_-)r_0\,dt.
\end{equation*}
This equation can be straightforwardly integrated and the result is
\begin{equation}
\left(\frac{c(t)-c_3}{c(0)-c_3}\right)^{-\alpha_3}
\left(\frac{c(t)-c_+}{c(0)-c_+}\right)^{-\alpha_+}
\left(\frac{c(t)-c_-}{c(0)-c_-}\right)^{-\alpha_-}= e^{3(c_+-c_-)r_0t/2}~,
\end{equation}
where
\begin{equation*}
\alpha_\pm=\frac{1}{c_3-c_\pm}~,\qquad \alpha_3=\alpha_--\alpha_+
= \frac{c_+-c_-}{(c_3-c_+)(c_3-c_-)}~.
\end{equation*}

The main feature of this result is that the approach to the reactive steady
state $c(t)\to c_-$ is still exponential in time, but the decay time can become
quite long when the parameter $x_0\approx \frac{1}{9}$.  For example for
$x_0\alt\frac{1}{9}$, the asymptotic decay of the minority density is
\begin{equation}
c(t)-c_- \sim  \exp\left[- \frac{3r_0}{2}(c_+-c_-)(c_--c_3)\, t\right].
\end{equation}
Thus as $x\to\frac{1}{9}$ where $c_-$ and $c_3$ approach each other, the
approach to the fixed point becomes very slow.  On the other hand, for
$x\agt\frac{1}{9}$, the densities all decay to their common value of
$\frac{1}{3}$ as
\begin{equation}
c(t)-c_3 \sim  \exp\left[- \frac{3r_0}{2}(c_3-c_+)(c_3-c_-)\, t\right].
\end{equation}

\section{Influence of stochastic fluctuations}

In a finite population, the stochasticity of the dynamics allows the system
to eventually escape the basin of attraction of a fixed point.  This escape
corresponds to a minority party becoming one of the top two parties on a slow
time scale.  To understand this escape dynamics, we formulate a discrete
version of the strategic voter model for a fixed population $N$, with $N_A$
agents of species $A$, $N_B$ of species $B$, $N_C$ of species $C$, with
$N=N_A+N_B+N_C$.  This population evolves by the reactions:
\begin{eqnarray*}
&A \stackrel{T}{\longrightarrow}
\begin{cases}
B\\
C
\end{cases}\!\!\!\!\!,
\quad
B \stackrel{T}{\longrightarrow}
\begin{cases}
A\\
C
\end{cases}\!\!\!\!\!,
 \quad
C \stackrel{T}{\longrightarrow}
\begin{cases}
A\\
B
\end{cases}\!\!\!\!\!,
\\
&~~~~~AB\stackrel{r}{\longrightarrow}
\begin{cases}
 AA \\
BB
\end{cases}\!\!\!\!\!,
 \quad  
AC \stackrel{r}{\longrightarrow} AA\,,  \qquad
BC \stackrel{r}{\longrightarrow}BB\,,
\end{eqnarray*}
for the case where $C$ is in the minority.  Corresponding rules apply when
$A$ and $B$ are in the minority. 

\begin{figure}[ht]
\centerline{\includegraphics*[width=0.355\textwidth]{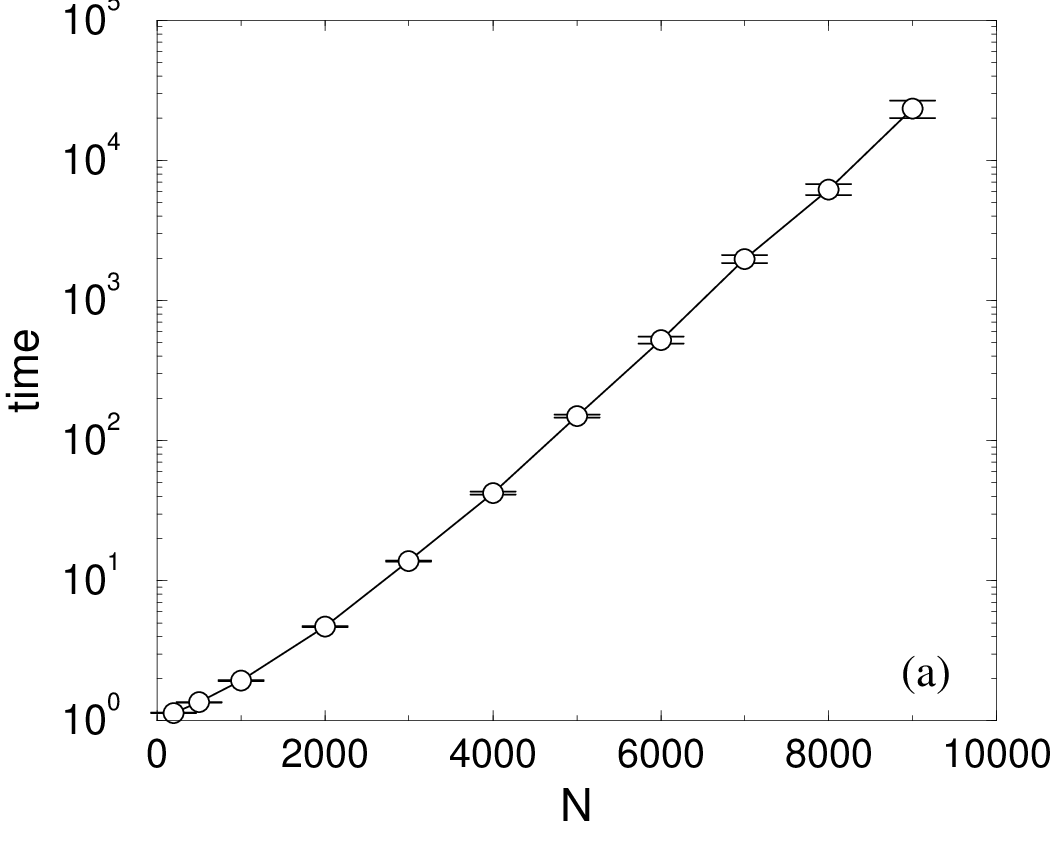}
\hskip 0.75in\includegraphics*[width=0.355\textwidth]{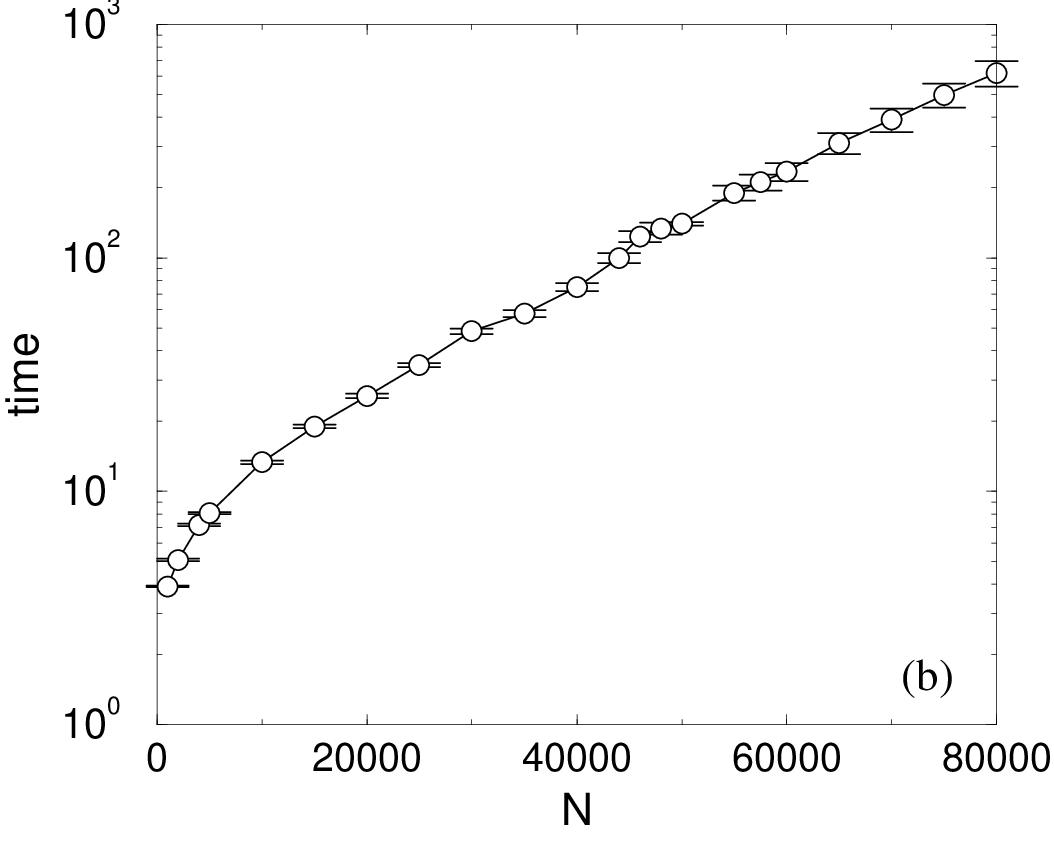}}
\caption{Average time span for a single party to have minority status versus
  population $N$ for strategic voting with: (a) constant strategic bias with
  $x=2$ and (b) state-dependent strategic bias with $x_0=0.104$; this latter
  value is chosen to give the same steady-state minority densities for both
  models.  Both time dependences are consistent with exponential growth in
  $N$.}
  \label{t-vs-N}
\end{figure}

The stochastic dynamics of the strategic voter model may be analytically
described by a master equation whose finite-size
expansion~\cite{Gardiner,vanKampen} leads to a Fokker-Planck equation.  The
fixed points in the rate equation then correspond, for finite $N$, to an
effective attractor for the probability distribution.  Because of finite-size
fluctuations, the system can escape such a potential well and thereby move
between different attractors.  This barrier crossing corresponds to a change
in the identity of the minority party.  We can investigate this barrier
crossing by rephrasing it as a first-passage
problem~\cite{Gardiner,vanKampen,Redner}.  Following a standard calculation
(as given, for example, in Ref.~\cite{RMF}), we find that changes in the
identity of the minority party occur on a time scale that grows exponentially
in $N$~\cite{InPrep}, much longer than the time scale of fluctuations between
the majority and second-place parties.

We have tested this prediction by numerical simulations.  The time span over
which a given state is in the minority as a function of the total population
$N$ does appear to grow exponentially with $N$ for both constant and
state-dependent strategic bias, but with a much smaller amplitude for the
latter case (Fig.~\ref{t-vs-N}).  For the more realistic case of a
state-dependent strategic bias, simulations of single realizations exhibit
changes in minority status on a time scale that can be tuned to be in a
similar range as that of election data (Fig.~\ref{elections}).

\section{Discussion}

We proposed a strategic voter model to describe the feature whereby two major
parties dominate in certain parliamentary democracies, while a third party
remains in the minority.  Our model is based on a strategic bias that
inherently disfavors the minority species.  We studied two variants of
strategic voting in which the strategic bias is either fixed or state
dependent.  For fixed bias, the minority species is doomed to remain
eternally in the minority in the rate equation limit.  For a state-dependent
bias, a bifurcation occurs between a stable minority and equal densities of
the three parties as the strength of the strategic bias varies.

It is worth mentioning that our three-choice voting model has an important
distinction with predator-prey models that exhibit "competitive exclusion".
This feature --- known as Gause's Law \cite{G} --- states that two species
that compete for the same resources cannot stably coexist if other ecological
factors are constant.  In our model, if we consider the opinions as different
species, the dynamics allow for spontaneous mutations and additionally the
interactions are not constant but state dependent.  These features allow for
a stable coexistence of all species.

For finite populations, stochastic fluctuations allow the minority party to
escape its status on a time scale that grows exponentially with population
size.  The concomitant slow oscillations in the identity of the minority
party can be readily seen in simulations, and this time evolution seems to
roughly mirror real election data, especially if the latter could be extended
over a much longer time scale.  However, to quantitatively match the election
data, it would be necessary to have a strategic bias that is still weaker
than our state-dependent strategic voter model to reproduce both the density
difference between majority and minority and the time scale for a change in
minority status.

\vspace{0.2cm}

{\bf Acknowledgments:} SR thanks the hospitality of the Complexity Science
Centre at the University of Warwick, where this work began.  We gratefully
acknowledge financial support from NSF grant DMR0535503 (DV and SR), and the
Swiss National Science Foundation through the Advanced Fellowship
PA002-119487 (MM).  


\begin{thebibliography}{99}
\bibitem{sc} We ignore Canadian political parties that had only regional
  strength, such as the Social Credit party.
\end{thebibliography}
\end{document}